# Habitability Models for Astrobiology

Abel Méndez, Planetary Habitability Laboratory, University of Puerto Rico at Arecibo, Puerto Rico, USA
Edgard G. Rivera-Valentín, Lunar and Planetary Institute, USRA, Houston, Texas, USA
Dirk Schulze-Makuch, Center for Astronomy and Astrophysics, Technische Universität Berlin, Berlin, Germany; German Research Centre for Geosciences, Section Geomicrobiology, Potsdam, Germany; Leibniz-Institute of Freshwater Ecology and Inland Fisheries, Stechlin, Germany.
Justin Filiberto, Lunar and Planetary Institute, USRA, Houston, Texas, USA
Ramses M. Ramírez, University of Central Florida, Department of Physics, Orlando, Florida, USA; Space Science Institute, Boulder, Colorado, USA.
Tana E. Wood, USDA Forest Service International Institute of Tropical Forestry, San Juan, Puerto Rico, USA
Alfonso Dávila, NASA Ames Research Center, Moffett Field, California, USA
Chris McKay, NASA Ames Research Center, Moffett Field, California, USA
Kevin N. Ortiz Ceballos, Planetary Habitability Laboratory, University of Puerto Rico at Arecibo, Puerto Rico, USA
Marcos Jusino-Maldonado, Planetary Habitability Laboratory, University of Puerto Rico at Arecibo, Puerto Rico, USA
Nicole J. Torres-Santiago, Planetary Habitability Laboratory, University of Puerto Rico at Arecibo, Puerto Rico, USA
Guillermo Nery, Planetary Habitability Laboratory, University of Puerto Rico at Arecibo, Puerto Rico, USA
René Heller, Max Planck Institute for Solar System Research; Institute for Astrophysics, University of Göttingen, Germany
Paul K. Byrne, North Carolina State University, Raleigh, North Carolina, USA
Michael J. Malaska, Jet Propulsion Laboratory, California Institute of Technology, Pasadena, California, USA
Erica Nathan, Department of Earth, Environmental and Planetary Sciences, Brown University, Providence, Rhode Island, USA
Marta Filipa Simões, State Key Laboratory of Lunar and Planetary Sciences, Macau University of Science and Technology, Taipa, Macau SAR, China
André Antunes, State Key Laboratory of Lunar and Planetary Sciences, Macau University of Science and Technology, Taipa, Macau SAR, China
Jesús Martínez-Frías, Instituto de Geociencias (CSIC-UCM), Madrid, Spain
Ludmila Carone, Max Planck Institute for Astronomy, Heidelberg, Germany
Noam R. Izenberg, Johns Hopkins Applied Physics Laboratory, Laurel, Maryland, USA
Dimitra Atri, Center for Space Science, New York University Abu Dhabi, United Arab Emirates
Humberto Itic Carvajal Chitty, Universidad Simón Bolívar, Caracas, Venezuela
Priscilla Nowajewski-Barra, Fundación Ciencias Planetarias, Santiago, Chile
Frances Rivera-Hernández, Georgia Institute of Technology, Atlanta, Georgia, USA
Corine Brown, Macquarie University, Sydney, Australia
Kennda Lynch, Lunar and Planetary Institute, USRA, Houston, Texas, USA
David Catling, University of Washington, Seattle, Washington, USA
Jorge I. Zuluaga, Institute of Physics / FCEN - Universidad de Antioquia, Medellín, Colombia
Juan F. Salazar, GIGA, Escuela Ambiental, Facultad de Ingeniería, Universidad de Antioquia, Medellín, Colombia
Howard Chen, Northwestern University, Evanston, Illinois, USA
Grizelle González, USDA Forest Service International Institute of Tropical Forestry, San Juan, Puerto Rico, USA
Madhu Kashyap Jagadeesh, Jyoti Nivas College, Bengaluru, India
Jacob Haqq-Misra, Blue Marble Space Institute of Science, Seattle, Washington, USA

Habitability has been generally defined as the capability of an environment to support life. Ecologists have been using Habitat Suitability Models (HSMs) for more than four decades to study the habitability of Earth from local to global scales. Astrobiologists have been proposing different habitability models for some time, with little integration and consistency among them, being different in function to those used by ecologists. Habitability models are not only used to determine if environments are habitable or not, but they also are used to characterize what key factors are responsible for the gradual transition from low to high habitability states. Here we review and compare some of the different models used by ecologists and astrobiologists and suggest how they could be integrated into new habitability standards. Such standards will help to improve the comparison and characterization of potentially habitable environments, prioritize target selections, and study correlations between habitability and biosignatures. Habitability models are the foundation of planetary habitability science and the synergy between ecologists and astrobiologists is necessary to expand our understanding of the habitability of Earth, the Solar System, and extrasolar planets.

# 1.   Introduction

Life on Earth is not equally distributed. There is a measurable gradient in the abundance and diversity of life, both spatially (*e.g.*, from deserts to rain forests) and temporally (e.g., from seasonal to geological time-scales). Our planet has also experienced global environmental changes from the Archean to the Anthropocene, which further conditioned life to a broad range of conditions. In general, *a habitable environment is a spatial region that might support some form of life* (Farmer, 2018), *albeit not necessarily one with life* (Cockell et al., 2012). One of the biggest problems in astrobiology is how to define and measure the habitability not only of terrestrial environments but also of planetary environments, from the Solar System to extrasolar planets (also known as *exoplanets*). The word *habitability* literally means the *quality of habitat* (the suffix *-ity* denotes a quality, state, or condition). Astrobiologists have been constructing different general definitions of habitability, not necessarily consistent with one another (*e.g.*, Shock & Holland, 2007; Hoehler, 2007; Cardenas *et al*., 2014; Cockell *et al*., 2016; Cárdenas *et al*., 2019; Heller, 2020). Other more specific habitability definitions, such as the canonical Habitable Zone (*i.e.*, the orbital region in which liquid water could exist on the surface of a planet), are used in exoplanet science (Kasting et al., 1993). Ecologists developed a standardized system for defining and measuring habitability in the early 1980s; however, this is seldom utilized in the astrobiology community (USFWS, 1980).

The popular term habitability is formally known as *habitat suitability* in biology. Ecologists before the 1980s were using different and conflicting measures of habitability, a situation not much different than today for astrobiologists. The U.S. Fish and Wildlife Service (USFWS) decided to solve this problem with the development of the Habitat Evaluation Procedures (HEP) standards in 1974 for use in impact assessment and project planning (USFWS, 1980). These procedures include the development and application of Habitat Suitability Models (HSM) (Hirzel & Lay, 2008). Other names for these models are Ecological Niche Models (ENM), Species Distribution Models (SDM), Habitat Distribution Models (HDM), Climate Envelope Models (CEM), Resource Selection Functions (RSF), and many other minor variants (Guisan *et al*., 2017). These multivariate statistical models are widely used by ecologists today to quantify species-environment relationships with data obtained from both ground and satellite observations. HSMs integrate concepts as needed from ecophysiology, niche theory, population dynamics, macroecology, biogeography, and the metabolic theory of ecology.

Astrobiologists have largely not utilized HSMs for at least three reasons. First is the naming: habitability is a common word in the earth and planetary sciences, but it is not generally used by biologists. Thus, a quick review of the scientific literature shows no definition of this concept in biological terms. The second reason is the specialization: HSM is a specialized topic of theoretical ecology, which is not highly represented in the astrobiology community. The third is applicability: HSMs are mostly used to study the distribution of specific wild animals and plants, not microbial communities or ecosystems in general (generally the focus of astrobiological studies), so it may not seem readily applicable to the field of astrobiology, but this is changing (*e.g*., Treseder *et al*., 2012). For example, microorganisms (bacteria, fungi, and other unicellular life) exhibit endosymbiotic relationships with animals and plants, and also play a key role in their survival. Thus, anything that can be said about habitability at the macroscopic level is tightly coupled to habitability at the microscopic level. Indeed, potential methods for incorporating microbial ecology into ecosystem models are discussed in Treseder *et al.* (2012). In one way, the mathematical framework behind HSM is easier to apply to microbial communities than animals because the spatial interactions of animals (*e.g.*, predation) tend to be much more complex. However, microbial life is not easy to quantify in free-living populations and it is thus harder to validate the HSMs with them, although molecular methods are changing this (Douglas, 2018).

The definition and core framework of HSMs can be extended from the Earth to other planetary environments. However, the astrobiology field does not have the luxury of validating HSMs with the presence of life unless when applied to environments on Earth (*e.g.*, extreme environments). Thus, known ecophysiology models are used instead to predict the occurrence, distribution, and abundance of putative life in any planetary environment. Attempting to measure the habitability of a system without knowing all the environmental factors controlling it may seem like an impossible task. However, even on Earth, this problem can be approached by selecting a minimum set of relevant factors to simplify the characterization of the systems. While the objective can be to establish if a system is habitable, it can also be simply to explore how the selected environmental variables contribute to the habitability of the system. Usually, a library of habitability metrics is created for each environment or lifeform under consideration, with each metric dependent on the species, scales, or environmental factors under consideration. In a fundamental sense, the only way to really know if a place is habitable or not is to find (or put) life on it (Zuluaga *et al*., 2014; Chopra & Lineweaver, 2016). It is nearly impossible, nor is it desirable, to include all factors affecting habitability in a model, even for environments on Earth. Thus, the objective of habitability models is to understand the contributions of a *finite* set of variables toward the *potential* to support a specific species or community (e.g., primary producers, organisms that use abiotic sources of energy) (Guisan *et al*., 2017). So, even if we do not know or do not include all the relevant factors, we can consider the effects of those we do know.

Here, we recommend adapting and expanding the ecologists' nearly four decades of experience in modeling habitability on Earth to astrobiological studies. These models can be used to characterize the spatial and temporal distribution of habitable environments, identify regions of interest in the search for life, and, eventually, explore correlations between habitability and biosignatures. For example, such models would help to test the hypothesis that biosignatures (or *biomarkers*) are positively correlated with proxy indicators of geologically habitable environments (or *geomarkers*); *i.e.,* there is life whenever there are habitable environments on Earth (Martinez-Frias *et al*., 2007). We also note that the concept of biosignatures encompasses any detectable signature of life or its byproducts on a planet's atmosphere, which includes possible signatures of planetary-scale technology, known as *technosignatures* (Wright & Gelino, 2018). Measurements by past and future astrobiology related observations (e.g., from ground, telescopes, or planetary missions) can be combined into a standard library of habitability models. Results from different observations can then be compared, even using different measurements, since, through the use of HSMs, their results can be mapped to the same standard scale (*e.g.*, zero for worst and one for best regions). A Habitability Readiness Analysis (HRA) could be developed for any observation campaign to determine how its current instruments could be used, or what new instruments should be added, for habitability measurements in the spatial and temporal habitability scales of interest. Furthermore, it might also be possible to develop new instruments for direct habitability measurements.

This review addresses many of the misconceptions about habitability and stresses the need for better integration between the habitability models used by ecologists and astrobiologists. This is not a review of those factors affecting habitability, which are discussed elsewhere (e.g., National Research Council, 2007; Des Marais *et al*., 2008; McKay, 2014; Hendrix *et al*., 2018), but about the multivariate models that integrate these factors. Section 2 presents an overview of current ecology models with an emphasis on the Habitat Suitability Models. Section 3 discusses some examples of how habitability is currently implemented in the astrobiology field. Section 4 describes our recommendations on how to adapt and expand the ecology models to the astrobiology field. Section 5 presents important science questions that could be answered from habitability models. Finally, Section 7 presents our concluding remarks.

## 2.   Habitability in Ecology

Habitat Suitability Models (HSMs) are widely used in ecology to study the habitability of environments, many times under different definitions: species distribution models (SDMs) or environmental niche models (ENMs) (Kuhn *et al.*, 2016; Guisan *et al.*, 2017). An important step in the construction of HSMs is the selection of spatially explicit environmental variables at the right resolution to determine a species' preferred environment (*i.e.*, its niche) as close to its ecophysiological requirements as possible. Environmental variables (such as edaphic, from the Greek noun "edaphos" meaning ground factors – defined as any chemical, physical and biological properties of the soil) can exert complex direct or indirect effects on species (*e.g.*, Oren, 1999; Oren, 2001; Fierer *et al*., 2007; Lauber *et al*., 2008; Rajakaruna & Boyd, 2008; Allison & Martiny, 2008; Fierer *et al*., 2012). These variables are ideally chosen to reflect the three main types of influence on a species: (1) regulators or limiting factors, defined as factors controlling a species' metabolism (*e.g.*, physical-chemical conditions such as temperature and salinity); (2) disturbances, defined as any perturbations affecting environmental systems; and (3) resources, defined as all compounds that can be consumed by organisms (*e.g.*, nutrients). There are many other variables that exert an indirect, rather than a direct, effect on species distribution. The construction of HSMs follows five general steps: (1) conceptualization; (2) data preparation; (3) model calibration; (4) model evaluation; and (5) spatial predictions (Guisan *et al*., 2017).

One of the main HSM tools is the *Habitat Suitability Index* (HSI), which provides one way of quantifying the capacity of a given habitat to support a selected species. An index is the ratio of a value of interest divided by a standard of comparison. The value of interest is an estimate or measure of the quality of habitat conditions for a species in the studied environment, and the standard of comparison is the corresponding value for the optimum habitat conditions for the same evaluated species. An HSI of zero (minimum value) represents a totally unsuitable habitat, and a maximum value of one represents an optimum habitat. In developing an HSI we should obtain a direct and linear relationship between the HSI value and the carrying capacity of the environment for the species under consideration (USFWS, 1980). The functions describing the species distribution or abundance along each environmental variable in an HSM are called *species response curves* (Austin & Gaywood, 1994). These curves, when plotted, can vary from simple box-like envelopes resulting in binary indices to more gradual and complex responses resulting in continuous indices (Figure 1).

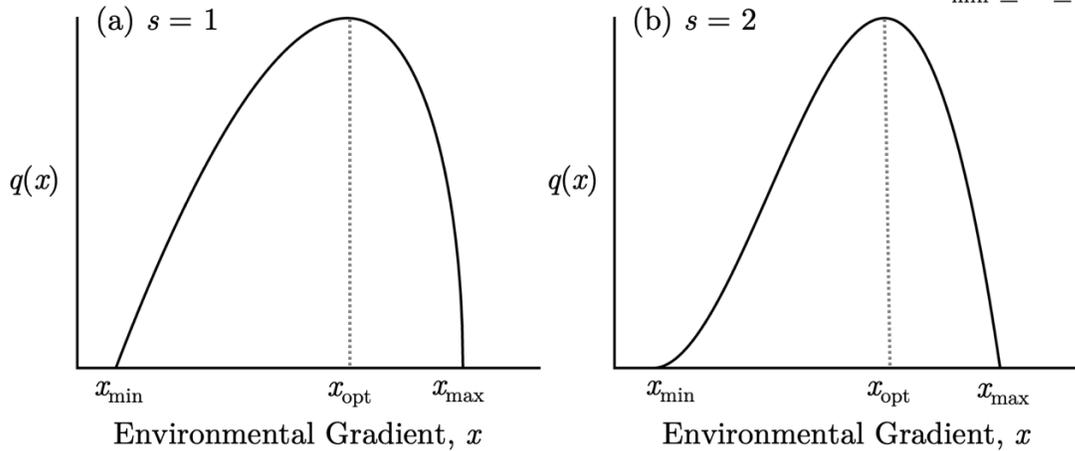

Normalized Biological Performance, $q(x) = \left[ \left( \dfrac{x - x_{\min}}{x_{\text{opt}} - x_{\min}} \right) \left( \dfrac{x_{\max} - x}{x_{\max} - x_{\text{opt}}} \right)^{\frac{x_{\max} - x_{\text{opt}}}{x_{\text{opt}} - x_{\min}}} \right]^s$

$x_{\min} \leq x \leq x_{\max}$

(a) $s = 1$   (b) $s = 2$

$q(x)$ model adapted from Yin *et al.* (1995)

**Figure 1.** Typical shapes of ecophysiological or species response curves along an environmental gradient (*e.g.*, temperature) for the biological performance (*e.g.*, relative growth rate) of plants (a) and microbial life (b) modeled after Yin *et al.* (1995). Species perform best near a physiological optimum and decrease asymmetrically near the extremes. The response curves of different species vary in shape and amplitude, and are subject to biological evolution (Lenton & Lovelock, 2000). Responses of multiple variables or species could be combined with different aggregations statistics (*e.g.*, arithmetic or geometric means).

Carrying capacity is generally defined as the maximum supported population density in equilibrium. More precisely, carrying capacity is the user-specified quality biomass of a particular species for which a particular area will supply all energetic and physiological requirements over a long, but specified, period (Giles, 1978). Since habitability could be taken as proportional to carrying capacity, as defined by the HSI, it is then related to the fraction of mass (*e.g.*, nutrients) and energy (*e.g.*, light) available or usable by a particular species or community from the environment. A common and difficult problem for HSIs is how to combine the effect of many environmental variables into a single index. In theoretical ecology, the solutions are known as aggregation methods. These methods can combine the variables using arithmetic, geometric, or harmonic means, among others. The general rule is to keep the index proportional to carrying capacity and correlated with the presence and absence of the species of interest in the environment. Occurrences or presence probabilities are generally simpler to combine as products. Ecophysiological response curves often involve the fitting of standard statistical models to ecological data using simple (multiple) regression, Generalised Linear Models (GLM), Generalised Least Squares (GLS), or Generalised Additive Models (GAM), among others.

The usual approach is to create a library of HSI models for all species (or communities) and environments under consideration, each with its own particular limitations (Brooks, 1997; Roloff & Kernohan, 1999). These models are easy to compare and combine since they use the same uniform scale (*e.g.*, a value between zero and one, proportional to the carrying capacity). Thus, each HSI is only applicable to a specific type of life and habitat as a function of a finite set of environmental variables within selected spatial and temporal scales. There are many other tools of the HSM that can be used to characterize species or their environment. For example, *similarity indices* are usually simpler to construct than an HSI and can be used for quick comparisons between a set of biological or physical properties (*e.g.*, diversity) (Boyle *et al.*, 1990). Similarity indices are also used in many other applications such as pattern recognition and machine learning (*e.g.*, Cheng *et al.*, 2011).

## 3. Habitability in Astrobiology

Astrobiologists have proposed many habitability models or indices for Earth, the Solar System, and extrasolar bodies in the last decade (*e.g.*, Stoker *et al.*, 2010; Schulze-Makuch *et al.*, 2011; Armstrong *et al.*, 2014; Irwin *et al.*, 2014; Barnes *et al.*, 2015; Silva *et al.*, 2017; Kashyap Jagadeesh *et al.*, 2017; Rodríguez-López *et al.*, 2019, Seales & Lenardic, 2020). There are some specific universal biological quantities that can be used as proxies for habitability such as carrying capacity, growth rate, metabolic rate, productivity, or the presence of some requirements of life, or even genetic diversity (Heller 2020). There is also an ongoing debate as to whether any concept of habitability needs to be binary (yes/no) in nature (Cockell *et al.* 2019), continuous (Heller, 2020), or probabilistic (Catling *et al.*, 2018). While a binary interpretation of habitability only allows a given planet to be habitable (for a given species) or not, a continuous model also allows for the possibility of a world (planet or moon) to be even more habitable than Earth, that is, to be superhabitable (Heller & Armstrong, 2014; Schulze-Makuch *et al.*, 2020). Constructing a direct measure of habitability requires knowing how the environment affects one of the biological quantities for some species or community. We do not need to specifically estimate these quantities, but only to know how the environment proportionally affects them. For example, we know how temperature affects the productivity of primary producers such as plants and phytoplankton. Most require temperatures between 0° and 50° C, but such producers do better (*i.e.*, have the highest productivity) near 25° C (Silva et al., 2017). Their 'thermal habitability function' looks like a bell-shaped curve centered at their optimum productivity temperature. Direct measures of habitability are also better represented as a fraction from zero to one.

Biological productivity (the dry or carbon biomass produced over space and time) is one of the best habitability proxies for astrobiology since it is easy to estimate for many ecosystems, via ground or satellite observations. The *Miami Model* was the first global-scale empirical model to give fair estimates of terrestrial net primary productivity (NPP, the rate of fixed photosynthetic carbon minus the carbon used by autotrophic respiration) (Zaks *et al.*, 2007). This simple model only uses two measurements, annual mean surface temperature and precipitation, to successfully infer the global distribution of vegetation (Adams *et al*., 2004). One important limitation of this type of model is that climate variables such as precipitation not only affect but are also affected by vegetation. There is increasing evidence, for example, that tropical forests have strong impacts on cloud base heights (Van Beusekom *et al.,* 2017) and precipitation patterns on Earth (Molina *et al*., 2019). Today, many complex biogeochemical models and satellite observations (e.g., NASA's TERRA, AQUA, and Soumi NPP models) are combined to estimate local to global NPPs (Cramer *et al*., 1999; Ito, 2011). These satellite products are being used to create habitability indices to monitor terrestrial biodiversity now and through climate change (*e.g.,* Pan *et al*., 2010; Radeloff *et al*., 2019). Therefore, the NPP is also a measure of global terrestrial health or habitability since primary producers are the basis of the food chain.

Most habitability models are limited to indirect measures of habitability due to a lack of information. This is especially true for exoplanets. For example, the occurrence of Earth-sized planets in the Habitable Zone of stars (termed the *Eta-Earth* value) can be considered a continuous indirect measure of stellar habitability (*i.e.,* the suitability of stars for habitable planets). The Habitable Zone, the region around a star where an Earth-like planet could maintain surface liquid water, is generally considered to be a binary indirect measure of planetary habitability (Kasting *et al*., 1993), although others have argued that it should be regarded as a probability density function (Zsom, 2015; Catling *et al*., 2018). While the location of the Habitable Zone depends on stellar type, its extension also greatly depends on the assumed atmospheric composition (*e.g.*, Heng 2016). Furthermore, atmospheric dynamics effectively work to homogenize differential heating of the surface, creating a short-term response on the planet's global temperature. This differential heating is a result of the planet's obliquity, which governs the latitudinal distribution of incoming stellar radiation (Spiegel *et al*., 2009; Nowajewski *et al*., 2018). The Habitable Zone boundaries themselves also evolve over time. This has major implications for water delivery, water retention, and oxygen buildup on potentially habitable planets (*e.g.*, Ramirez & Kaltenegger 2014; Luger & Barnes 2015).

The Habitable Zone can be defined in terms of either the planet's distance from the star, its incoming stellar flux, or its global equilibrium temperature. When using the equilibrium temperature definition, the extension of the Habitable Zone depends on the planet's orbital forcings, particularly eccentricity and obliquity. For example, when orbital eccentricity increases, the average equilibrium temperature decreases, thus extending the size of the Habitable Zone (Méndez & Rivera-Valentín, 2017). Similarly, higher fixed obliquity and/or rapid changes in obliquity values result in higher average equilibrium temperatures, which also result in extending the outer edge of the Habitable Zone (Armstrong *et al*., 2014). Furthermore, when using the equilibrium temperature definition, the extension of the Habitable Zone depends ultimately on the planet's energy balance. On Earth, the global energy balance is a result of the complex interaction between physical and biological processes. Biota affect the global energy balance in manifold ways including direct effects on surface albedo and latent heat fluxes (*e.g*., transpiration) (Jasechko *et al*., 2013; Duveiller *et al*., 2018). Tidal heating from the newly formed and nearby Moon might have played a role early in Earth's history (Heller *et al.,* 2020) but is irrelevant today. On Earth-sized planets in the Habitable Zones around M dwarf stars, tidal heating can have a strong effect on the planetary energy budget, potentially making some parts of the Habitable Zone uninhabitable (Barnes *et al.*, 2009).

The Earth Similarity Index (ESI), inspired by the diversity similarity indices used in ecology to compare populations (Boyle *et al.*, 1990), is a measure of Earth-likeness for a selected set of planetary parameters (Schulze-Makuch *et al.*, 2011). Future observational constraints of Earth-similar atmospheric constituents (*i.e.*, $N_2$, $CO_2$, $H_2O$) could improve our handle on this and similar metrics. For instance, 3D global climate models indicate that spectral features of water vapor on close-in terrestrial exoplanetary atmospheres may be detectable by the *James Webb Space Telescope* (Kopporapu *et al.*, 2017; Chen *et al.*, 2019), depending on the presence of clouds (Komacek *et al.*, 2020). Even though the presence of water vapor in the atmospheres of terrestrial exoplanets can indicate habitability, it is necessary to perform exhaustive work to determine which species could survive under conditions of extreme humidity. For example, mammals are not capable of surviving hyperthermia produced under high air temperatures and high humidity conditions, so planets with extreme differential heating between latitudes may be uninhabitable for them, despite having liquid water on their surface (Nowajewski *et al.*, 2018). Furthermore, many animals (including all mammals) may not survive in atmospheres with $CO_2$ ($N_2$) pressures exceeding ~0.1 bars (Ramirez, 2020).

The current Habitable Zone paradigm is misunderstood by many people — the public, the press, as well as other scientists — but as with all habitability models, it has a specific application and is neither incorrect nor useless for neglecting the subsurface oceans in the outer Solar System, Venusian clouds, or other environments far from Earth-like conditions. The Habitable Zone does not tell us if planets are habitable (or even if there are planets there) but it shows the impact of a few important variables on planetary habitability. The concept of a Habitable Zone was developed to identify terrestrial exoplanet targets that could potentially host life. It was first proposed by Edward Maunder in 1913 (Maunder, 1913, Lorenz, 2020) in his book *Life on Other Planets*, with refining definitions later on (Huang, 1959; Hart, 1978; Kasting *et al.*, 1993; Underwood *et al.*, 2003; Selsis *et al.*, 2007; Kaltenegger & Sasselov, 2011; Kopparapu *et al.*, 2013, 2014; Ramirez & Kaltenegger 2017, 2018; Ramirez, 2018). The Habitable Zone can be defined as *the circumstellar region where standing bodies of liquid water could be stable on the surface of a terrestrial planet.* Here, the insistence on the presence of liquid surface water is based on the fact that all known examples of life on Earth require liquid water to exist. However, this definition is suitable *only* for remote observations of planets and does not consider any life which might exist within the subsurface. There is a reason for this: the search for life on exoplanets will rely on remote observations of atmospheres for the foreseeable future, lacking the luxury of in-situ measurements used in solar system planetary science. Therefore, identifying water in the atmosphere of planets (in addition to other biosignature-relevant gases) is the only way to narrow down potential life-hosting targets, since subsurface life deep in the interior may not be able to modify the atmospheres of planets enough to be detectable remotely.

The abundance of liquid water in a planetary environment itself may be inherently unstable (Gorshkov et al., 2004), which leads to questions about the role of life in the definition of habitability itself (Zuluaga *et al.*, 2014). A disequilibrium may be one the most conspicuous signatures of a habitable and inhabited planet (Lovelock, 1965, 1975; Kleidon, 2012; Krissansen-Totton *et al.*, 2016); consider, for example, the composition of Earth's atmosphere (Lenton, 1998). One common problem with some, if not all, biological models is that they assume that the full set of physical characteristics of the environment, including climate, is *a boundary condition* for life, *i.e.* that biological systems depend on climate, but not the other way around. This premise is challenged by the fact that the observed state of the Earth system is the result of a complex and dynamic interaction between biological (*e.g.*, ecosystems) and physical (*e.g.*, climate) systems (Budyko, 1974; Gorshkov *et al.*, 2000; Kleidon, 2012; Zuluaga *et al.*, 2014). A critical question is how such a thermodynamically unstable state can be maintained during eons (*e.g.*, the span of Earth's life) despite variable (*e.g.*, solar luminosity) and sudden large (*e.g.*, asteroid impact) external forcings . The answer depends on the interactions between biological and physical systems on Earth. A planet might enter a habitable state (*i.e.*, allow for the presence of liquid water at its surface) at any given point in time simply through *chance occurrence* – a random change in planetary energy balance, for example. However, *long-term persistence* of a habitable state (*e.g.*, the persistent habitable state of Earth over the last 4 billion years, approximately) indicates the existence of natural regulation mechanisms (Walker *et al.*, 1981, Lenton, 1998; Gorshkov *et al.*, 2000; Kleidon & Lorenz, 2004, Salazar & Poveda, 2009).

## 4. Recommendations for Astrobiology

The astrobiology field is playing a critical role in our understanding of planetary habitability. The habitability of Earth, the Solar System, and exoplanets can be studied thanks to measurements taken with multiple ground, orbital, or remote sensors. At the same time, astrobiology-related missions can synergistically take advantage of the predictions of habitability models in their selection of potential exploration strategies, mission priorities, instrument design, and observations and experiments. Here we list four recommendations for the astrobiology community:

1. **Increase and widen the participation of more experts on habitat suitability models.** Ecologists are the experts in the ground-truth proven measurement of terrestrial habitability, yet they are seldom represented in the planetary and astrobiology community. New synergies between NASA and the national and international ecological societies, *e.g.*, the Ecological Society of America (ESA), Soil Ecology Society (SES), and the International Society for Microbial Ecology (ISME), should be established via, for example, a joint conference session at the Lunar and Planetary Science Conference, the Astrobiology Science Conference (AbSciCon), or the European Astrobiology Network Association (EANA). There should be worldwide participation to guarantee global standardization, which will stimulate exploration of the Solar System, promote use of the Solar System as a laboratory for expanding our current understanding of the habitability of Earth, and deepen our understanding of potentially habitable conditions elsewhere.

2. **Further terrestrial exploration.** Many Earth habitats (*e.g.* the cloud layer, stratosphere, deep ocean, deep ice, deep earth, or the mantle) are vastly under-explored biologically (*e.g.*, Lollar et al., 2019; DasSarma *et al.*, 2020). Astrobiologists should create stronger connections with the researchers working in these under-studied environments (*e.g.*, The Deep Carbon Observatory) to ensure that there is a cohesive understanding of the state-of-the-art science being learned and that continuing efforts to study these environments are supported. These field studies should provide new data to test the applicability of current habitability models with extreme environments, and thus identify a more diverse range of planetary and exoplanetary conditions. At the same time, unicellular life continually surprises us with new ways to survive and obtain energy from its environment (rock-eaters, electric currents, and even radioactivity) which shows us we need to be flexible in considering energy sources for habitability.

3. **Improve habitability models.** New habitability models should be developed and validated with field and laboratory experiments, including simulated extreme and planetary analog environments (*e.g.*, Taubner *et al.*, 2020). The main goal is to identify knowledge gaps. For example, new ecophysiological response curves for some organisms are necessary (*e.g.*, growth rate as a function of water activity – a measure of water available for biological reactions), especially in dynamic environments such as gradient-rich biotopes and higher complexity extreme environments (*i.e.*, those with multiple extremes, such as deep-sea brines). Also, there are insufficient models on microbial growth in near-surface dynamic environments (*e.g.*, as applicable to martian diurnal cycles). There is a growing body of literature about the manifold mechanisms through which life affects the Earth's climate system, including the global energy balance and atmospheric composition and dynamics. Advances in the understanding of climate-life interactions (e.g. Bonan and Doney, 2018) as well as climate-technology interactions (e.g., Frank et al. 2017) in the Earth System can provide new insights for habitability models.

4. **Develop a Habitability Standard.** Existing and future studies should specify how they assess habitability for each of their observations according to a shared habitability standard. For example, measurements of surface temperature and water vapor from landers or orbital missions could be converted into a simple habitability model. The advantage of a standard is that past and future missions could be compared to each other and their habitability assessments refined, while new habitability knowledge gaps could be identified. This dynamic standard should be evaluated and updated regularly by a diverse and multidisciplinary committee, for example during a Decadal Survey and/or mid-decade review. Currently, the closest concept to a standard is the specific language included in e.g. the NASA Astrobiology (Des Marais *et al.*, 2008) and Ocean Worlds (Hendrix *et al.*, 2018) roadmaps. These documents stress the need for habitability evaluations and missions (*e.g.,* Europa Clipper and Titan Dragonfly), yet only focus on the individual habitability requirements and not how to combine the net contribution of these factors. Furthermore, a habitability standard might eventually become the standard of other disciplines.

## 5. Science Questions

Each astrobiology-related project, mission, or instrument should anticipate and answer a series of basic scientific questions about the environment(s) to be studied as a core part of the planning process. The answers to these questions should be updated based on the results. To do so, it is important to define an environment of interest, both in space and time (termed a *quadrat* in ecology), and answer the following science questions as part of the initial analysis:

1. **What are the limiting factors?** Usually, there is a small set of main factors (*e.g.*, edaphic factors) that influence living organisms (*e.g.*, water, nutrients). These will be the first set of variables to be used for the construction of a habitability model, which will later be refined with more variables. For example, primary productivity is mainly driven by temperature, precipitation, and nutrients on land, and by temperature, and oxygen and nutrient concentrations in the oceans, among other factors. In general, these factors should be directly or indirectly related to the mass and energy of the environment (*e.g.*, Martiny *et al.*, 2006; Pikuta *et al.*, 2007; Williams & Hallsworth, 2009; Harrison *et al.*, 2013; McKay, 2014; Lynch & Neufeld, 2015; Tecon & Or, 2017). Another important factor is whether the organisms are able to come into contact with the limiting factors.

2. **What are the terrestrial and planetary analogs?** Identify at least one analog on Earth and one close planetary analog as the comparison standards (*i.e.*, model normalizations). For example, if studying a particular martian environment, select hyperarid terrestrial deserts and a martian analog based on the variables of interest. The cross-comparison of variants of similar types of environment (*e.g.*, salterns), as well as slightly different settings (*e.g.*, high salinity biotopes with different pH, temperature, or chemical conditions), could also prove useful. The subsurface oceans of Europa or Enceladus could be compared with deep seawater, hydrothermal systems, or deep-sea brines (Antunes *et al.*, 2020). Planetary atmospheres could be compared with high altitude or near-space regions. An analysis of similarities (*e.g.*, [ANOSIM](#)) could be used to formally select and compare these regions (Clarke, 1993).

3. **What is the habitability assessment?** The habitability of the region of interest is evaluated based on the selected environmental factors, and then compared with the selected Earth and planetary analogs, using a normalized scale from zero to one for simplicity. A library of habitability measures is usually constructed (*i.e.*, a habitability matrix), each for different considerations (*e.g.*, species). These inputs are then used to construct multivariate habitability maps (also known as *niche quantification* in ecology) for site selections. A common assumption is that habitability models are only used to determine if environments are habitable or not. Instead, they are used to characterize what key factors are responsible for the gradual transition from low to high habitability states. Therefore, a habitability threshold should also be defined to differentiate between habitable and non-habitable conditions.

4. **What is the potential biomass?** The upper limits of biomass can be predicted based on the fluxes of mass and energy available for life, and usually a very small fraction of the total mass and energy. For example, biomass could be estimated from the available metabolic energy using the Metabolic Theory of Ecology (van der Meer, 2006; Schramski *et al.*, 2015; Clarke, 2017). These upper limits are used in the sensitivity designs of life detection experiments. Available free energy from known disequilibria has been used to estimate an upper limit on the biomass in the subsurface of Mars and its value depends on uncertainties of the abundances of metabolic reactants and the assumed microbial basal power requirement (Sholes *et al.*, 2019).

5.  **What is the expected correlation between habitability and biosignatures?** The potential upper values of biomass can be converted to estimates of observable biosignatures or disequilibrium chemistry (Catling *et al*., 2018). Habitability and biosignatures are positively correlated on Earth but this might not necessarily be true for other planets. A zero or negative correlation could indicate an incorrect habitability model, or biological or technological processes unlike Earth (in other words, *life as we don't know it*). The habitability-biosignatures correlation is a fundamental problem of astrobiology, but non-detections are also important. For example, it will be profound to detect planetary regions determined to be habitable by Earth standards yet devoid of any detectable life. Such discoveries would place bounds on abiogenesis. Alternatively, if biosignatures are widespread in habitable regions but lack any plausible technosignatures, then this would suggest that the development of planetary-scale technology is a rare occurrence, even if life is common (Haqq-Misra *et al*. 2020).

## 6.   Conclusion

Habitability models are successful analysis tools for characterizing habitable environments on Earth. In this review, we compared some of the different models used by ecologists and astrobiologists and suggested how to integrate them into new habitability standards. These standards are relevant for any astrobiology-related observations, including the study of extreme environments on Earth, planetary missions, or exoplanets. Ecologists have been using these models for more than four decades to understand the distribution of terrestrial life at local to global scales (Section 2). Astrobiologists have been proposing different models for some time, with little integration and consistency between them and different in function to those used by biologists (Section 3). The astrobiology community should create habitability standards for observations and missions with astrobiology objectives, as the USFWS successfully did long ago for ecologists (Section 4). These standards are necessary to make sense of data from multiple observations, develop predictions for environmental niches that can be tested, and understand the extraterrestrial correlations between habitability and biosignatures (Section 5).

There is no need for the astrobiology community to reinvent the methods and tools used by ecologists. It is true that ecology methods are more capable than our limited planetary and astronomical data allow, but they also provide the basic language and framework to connect Earth and astrobiology science for decades to come. For example, there are many theoretical and computational tools used in ecology to quantify environments and their habitability, mostly known as habitat suitability models. See Guisan *et al*. (2017) for an extensive review of these models and Lortie *et al*. (2020) for a current review of the computational tools. Most of these tools are available as packages in the [R Computing Language](#) in the [Comprehensive R Archive Network (CRAN)](#) and [GitHub](#) (*e.g.,* [Environmetrics](#), [HSDM](#)). New, higher-resolution remote sensing instruments and exploration technologies will create better habitability maps from rover, lander, and orbiter data. Habitability models will eventually lead us to a better understanding of the potential for life in the Solar System and beyond, and perhaps even the factors that influence the development of life itself. *Habitability models are the foundation of planetary habitability science*. After all of our scientific and technological advances, we still need a stronger integration between biology, planetary sciences, and astronomy (Cockell, 2020).

**Acknowledgments**

This work was supported by a NASA Astrobiology Institute (NAI) workshop grant, the Planetary Habitability Laboratory (PHL), and the University of Puerto Rico at Arecibo (UPR Arecibo). Thanks to NASA Puerto Rico Space Grant Consortium and the Puerto Rico Louis Stokes Alliance For Minority Participation (PR-LSAMP) for supporting some of our students. Thanks to Ravi Kumar Kopparapu from the NASA Goddard Space Flight Center and James Kasting from Penn State for valuable comments. RH is supported by the German space agency (Deutsches Zentrum für Luft- und Raumfahrt) under PLATO Data Center grant 50OO1501. RMR is supported by the Earth-Life Science Institute and the National Institutes of Natural Sciences: Astrobiology Center (grant number JY310064). AA and MFS are funded by the Science and Technology Development Fund, Macau SAR. JHM gratefully acknowledges support from the NASA Exobiology program under grant 80NSSC20K0622. GG was supported by the NSF Luquillo Critical Zone Observatory (EAR-1331841) and the LTER program (DEB 1831952). JF acknowledges partial support from NASA PSTAR grant 80NSSC18K1686. The US Department of Agriculture (USDA) Forest Service's International Institute of Tropical Forestry (IITF) and UPR Río Piedras provided additional support. A portion of the research by MM was carried out at the Jet Propulsion Laboratory, California Institute of Technology, under a contract with the National Aeronautics and Space Administration (80NM0018D0004). This is LPI contribution number 2596. LPI is operated by USRA under a cooperative agreement with the Science Mission Directorate of the National Aeronautics and Space Administration.

## 7.   References

Adams, B., White, A., & Lenton, T. M. (2004). An analysis of some diverse approaches to modelling terrestrial net primary productivity. *Ecological Modelling*, *177*(3), 353–391. https://doi.org/10.1016/j.ecolmodel.2004.03.014

Allison, S. D., & Martiny, J. B. H. (2008). Resistance, resilience, and redundancy in microbial communities. *Proceedings of the National Academy of Sciences*, *105*(Supplement 1), 11512–11519. https://doi.org/10.1073/pnas.0801925105

Antunes, A., Olsson-Francis, K., & McGenity, T. J. (2020). Exploring Deep-Sea Brines as Potential Terrestrial Analogues of Oceans in the Icy Moons of the Outer Solar System. *Current Issues in Molecular Biology*, 123–162. https://doi.org/10.21775/cimb.038.123

Armstrong, J. c., Barnes, R., Domagal-Goldman, S., Breiner, J., Quinn, T. r., & Meadows, V. s. (2014). Effects of Extreme Obliquity Variations on the Habitability of Exoplanets. *Astrobiology*, *14*(4), 277–291. https://doi.org/10.1089/ast.2013.1129

Austin, M. P., & Gaywood, M. J. (1994). Current problems of environmental gradients and species response curves in relation to continuum theory. *Journal of Vegetation Science*, *5*(4), 473–482. https://doi.org/10.2307/3235973

Barnes, R., Jackson, B., Greenberg, R., & Raymond, S. N. (2009). TIDAL LIMITS TO PLANETARY HABITABILITY. *The Astrophysical Journal*, *700*(1), L30–L33. https://doi.org/10.1088/0004-637X/700/1/L30

Barnes, R., Meadows, V. S., & Evans, N. (2015). COMPARATIVE HABITABILITY OF TRANSITING EXOPLANETS. *The Astrophysical Journal*, *814*(2), 91. https://doi.org/10.1088/0004-637X/814/2/91

Bonan, G. B., & Doney, S. C. (2018). Climate, ecosystems, and planetary futures: The challenge to predict life in Earth system models. *Science*, *359*(6375). https://doi.org/10.1126/science.aam8328

Boyle, T. P., Smillie, G. M., Anderson, J. C., & Beeson, D. R. (1990). A Sensitivity Analysis of Nine Diversity and Seven Similarity Indices. *Research Journal of the Water Pollution Control Federation*, *62*(6), 749–762. JSTOR.

Brooks, R. P. (1997). Improving Habitat Suitability Index Models. *Wildlife Society Bulletin (1973-2006)*, *25*(1), 163–167. JSTOR.

Budyko (Ed.). (1974). *Climate and life* (English ed edition). Academic Press.

Cárdenas, R., Nodarse-Zulueta, R., Perez, N., Avila-Alonso, D., & Martin, O. (2019). On the Quantification of Habitability: Current Approaches. In R. Cárdenas, V. Mochalov, O. Parra, & O. Martin (Eds.), *Proceedings of the 2nd International Conference on BioGeoSciences* (pp. 1–8). Springer International Publishing. https://doi.org/10.1007/978-3-030-04233-2_1

Cardenas, R., Perez, N., Martinez-Frias, J., & Martin, O. (2014). On the Habitability of Aquaplanets. *Challenges*, *5*(2), 284–293. https://doi.org/10.3390/challe5020284

Catling, D. C., Krissansen-Totton, J., Kiang, N. Y., Crisp, D., Robinson, T. D., DasSarma, S., Rushby, A. J., Del Genio, A., Bains, W., & Domagal-Goldman, S. (2018). Exoplanet Biosignatures: A Framework for Their Assessment. *Astrobiology*, *18*(6), 709–738. https://doi.org/10.1089/ast.2017.1737

Chen, H., Wolf, E. T., Zhan, Z., & Horton, D. E. (2019). Habitability and Spectroscopic Observability of Warm M-dwarf Exoplanets Evaluated with a 3D Chemistry-Climate Model. *The Astrophysical Journal*, *886*(1), 16. https://doi.org/10.3847/1538-4357/ab4f7e

Cheng, F., Shen, J., Yu, Y., Li, W., Liu, G., Lee, P. W., & Tang, Y. (2011). In silico prediction of Tetrahymena pyriformis toxicity for diverse industrial chemicals with substructure pattern recognition and machine learning methods. *Chemosphere*, *82*(11), 1636–1643. https://doi.org/10.1016/j.chemosphere.2010.11.043

Chopra, A., & Lineweaver, C. H. (2016). The Case for a Gaian Bottleneck: The Biology of Habitability. *Astrobiology*, *16*(1), 7–22. https://doi.org/10.1089/ast.2015.1387

Clarke, A. (2017). The Metabolic Theory of Ecology. In *Principles of Thermal Ecology: Temperature, Energy, and Life*. Oxford University Press. https://www.oxfordscholarship.com/view/10.1093/oso/9780199551668.001.0001/oso-9780199551668-chapter-12

Clarke, K. R. (1993). Non-parametric multivariate analyses of changes in community structure. *Australian Journal of Ecology*, *18*(1), 117–143. https://doi.org/10.1111/j.1442-9993.1993.tb00438.x

Cockell, C. S., Balme, M., Bridges, J. C., Davila, A., & Schwenzer, S. P. (2012). Uninhabited habitats on Mars. *Icarus*, 217(1), 184–193. https://doi.org/10.1016/j.icarus.2011.10.025

Cockell, C. s., Bush, T., Bryce, C., Direito, S., Fox-Powell, M., Harrison, J. p., Lammer, H., Landenmark, H., Martin-Torres, J., Nicholson, N., Noack, L., O'Malley-James, J., Payler, S. j., Rushby, A., Samuels, T., Schwendner, P., Wadsworth, J., & Zorzano, M. p. (2016). Habitability: A Review. *Astrobiology*, *16*(1), 89–117. https://doi.org/10.1089/ast.2015.1295

Cockell, C. S., Stevens, A. H., & Prescott, R. (2019). Habitability is a binary property. *Nature Astronomy*, *3*(11), 956–957. https://doi.org/10.1038/s41550-019-0916-7

Cockell, C. S. (2020). Astronomy + biology. *Astronomy & Geophysics*, *61*(3), 3.28-3.32. https://doi.org/10.1093/astrogeo/ataa042

Cramer, W., Kicklighter, D. W., Bondeau, A., Iii, B. M., Churkina, G., Nemry, B., Ruimy, A., Schloss, A. L., & Intercomparison, T. P. O. T. P. N. M. (1999). Comparing global models of terrestrial net primary productivity (NPP): Overview and key results. *Global Change Biology*, *5*(S1), 1–15. https://doi.org/10.1046/j.1365-2486.1999.00009.x

DasSarma, P., Antunes, A., Simões, M. F., & DasSarma, S. (2020). Earth's Stratosphere and Microbial Life. *Current Issues in Molecular Biology*, 197–244. https://doi.org/10.21775/cimb.038.197

Des Marais, D. J., Nuth, J. A., Allamandola, L. J., Boss, A. P., Farmer, J. D., Hoehler, T. M., Jakosky, B. M., Meadows, V. S., Pohorille, A., Runnegar, B., & Spormann, A. M. (2008). The NASA Astrobiology Roadmap. *Astrobiology*, *8*(4), 715–730. https://doi.org/10.1089/ast.2008.0819

Douglas, A. E. (2018). What will it take to understand the ecology of symbiotic microorganisms? *Environmental Microbiology*, *20*(6), 1920–1924. https://doi.org/10.1111/1462-2920.14123

Duveiller, G., Hooker, J., & Cescatti, A. (2018). The mark of vegetation change on Earth's surface energy balance. *Nature Communications*, *9*(1), 679. https://doi.org/10.1038/s41467-017-02810-8

Farmer, J. D. (2018). Chapter 1—Habitability as a Tool in Astrobiological Exploration. In N. A. Cabrol & E. A. Grin (Eds.), *From Habitability to Life on Mars* (pp. 1–12). Elsevier. https://doi.org/10.1016/B978-0-12-809935-3.00002-5

Fierer, N., Bradford, M. A., & Jackson, R. B. (2007). Toward an Ecological Classification of Soil Bacteria. *Ecology*, *88*(6), 1354–1364. https://doi.org/10.1890/05-1839

Fierer, N., Lauber, C. L., Ramirez, K. S., Zaneveld, J., Bradford, M. A., & Knight, R. (2012). Comparative metagenomic, phylogenetic and physiological analyses of soil microbial communities across nitrogen gradients. *The ISME Journal*, *6*(5), 1007–1017. https://doi.org/10.1038/ismej.2011.159

Frank, A., Kleidon, A., & Alberti, M. (2017). Earth as a Hybrid Planet: The Anthropocene in an Evolutionary Astrobiological Context. *Anthropocene*, *19*, 13–21. https://doi.org/10.1016/j.ancene.2017.08.002

Giles, R. H. (1978). *Wildlife Management*. W. H. Freeman.

Gorshkov, V. G., Makarieva, A. M., & Gorshkov, V. V. (2004). Revising the fundamentals of ecological knowledge: The biota–environment interaction. *Ecological Complexity*, *1*(1), 17–36. https://doi.org/10.1016/j.ecocom.2003.09.002

Gorshkov, V., Makarieva, A. M., & Gorshkov, V. V. (2000). *Biotic Regulation of the Environment: Key Issues of Global Change*. Springer Science & Business Media.

Guisan, A., Thuiller, W., & Zimmermann, N. E. (2017). *Habitat Suitability and Distribution Models: With Applications in R*. Cambridge University Press.

Haqq-Misra, J., Kopparapu, R. K., & Schwieterman, E. (2020). Observational Constraints on the Great Filter. *Astrobiology*, *20*(5), 572–579. https://doi.org/10.1089/ast.2019.2154

Harrison, J. P., Gheeraert, N., Tsigelnitskiy, D., & Cockell, C. S. (2013). The limits for life under multiple extremes. *Trends in Microbiology*, *21*(4), 204–212. https://doi.org/10.1016/j.tim.2013.01.006

Hart, M. H. (1978). The evolution of the atmosphere of the earth. *Icarus*, *33*, 23–39. https://doi.org/10.1016/0019-1035(78)90021-0

Heller, R. (2020). Habitability is a continuous property of nature. *Nature Astronomy*, *4*(4), 294–295. https://doi.org/10.1038/s41550-020-1063-x

Heller, R., & Armstrong, J. (2014). Superhabitable Worlds. *Astrobiology*, *14*(1), 50–66. https://doi.org/10.1089/ast.2013.1088

Heller, R., Duda, J.-P., Winkler, M., Reitner, J., & Gizon, L. (2020). Habitability of the early Earth: Liquid water under a faint young Sun facilitated by strong tidal heating due to a nearby Moon. *ArXiv:2007.03423 [Astro-Ph]*. http://arxiv.org/abs/2007.03423

Hendrix, A. R., Hurford, T. A., Barge, L. M., Bland, M. T., Bowman, J. S., Brinckerhoff, W., Buratti, B. J., Cable, M. L., Castillo-Rogez, J., Collins, G. C., Diniega, S., German, C. R., Hayes, A. G., Hoehler, T., Hosseini, S., Howett, C. J. A., McEwen, A. S., Neish, C. D., Neveu, M., … Vance, S. D. (2018). The NASA Roadmap to Ocean Worlds. *Astrobiology*, *19*(1), 1–27. https://doi.org/10.1089/ast.2018.1955

Heng, K. (2016). The Imprecise Search for Extraterrestrial Habitability. *American Scientist*, *104*(3), 146. https://doi.org/10.1511/2016.120.146

Heng, K. (2017, February 6). *The Imprecise Search for Extraterrestrial Habitability*. American Scientist. https://www.americanscientist.org/article/the-imprecise-search-for-extraterrestrial-habitability

Hirzel, A. H., & Lay, G. L. (2008). Habitat suitability modelling and niche theory. *Journal of Applied Ecology*, *45*(5), 1372–1381. https://doi.org/10.1111/j.1365-2664.2008.01524.x

Hoehler, T. M. (2007). An Energy Balance Concept for Habitability. *Astrobiology*, *7*(6), 824–838. https://doi.org/10.1089/ast.2006.0095

Huang, S.-S. (1959). The Problem of Life in the Universe and the Mode of Star Formation. *Publications of the Astronomical Society of the Pacific*, *71*, 421. https://doi.org/10.1086/127417

Irwin, L. N., Méndez, A., Fairén, A. G., & Schulze-Makuch, D. (2014). Assessing the Possibility of Biological Complexity on Other Worlds, with an Estimate of the Occurrence of Complex Life in the Milky Way Galaxy. *Challenges*, *5*(1), 159–174. https://doi.org/10.3390/challe5010159

Ito, A. (2011). A historical meta-analysis of global terrestrial net primary productivity: Are estimates converging? *Global Change Biology*, *17*(10), 3161–3175. https://doi.org/10.1111/j.1365-2486.2011.02450.x

Jasechko, S., Sharp, Z. D., Gibson, J. J., Birks, S. J., Yi, Y., & Fawcett, P. J. (2013). Terrestrial water fluxes dominated by transpiration. *Nature*, *496*, 347–350. https://doi.org/10.1038/nature11983

Kaltenegger, L., & Sasselov, D. (2011). Exploring the Habitable Zone for Kepler Planetary Candidates. *The Astrophysical Journal Letters*, *736*, L25. https://doi.org/10.1088/2041-8205/736/2/L25
*Habitability Models* 16

Kashyap Jagadeesh, M., Gudennavar, S. B., Doshi, U., & Safonova, M. (2017). Indexing of exoplanets in search for potential habitability: Application to Mars-like worlds. *Astrophysics and Space Science*, *362*(8), 146. https://doi.org/10.1007/s10509-017-3131-y

Kasting, J. F., Whitmire, D. P., & Reynolds, R. T. (1993). Habitable Zones around Main Sequence Stars. *Icarus*, *101*(1), 108–128. https://doi.org/10.1006/icar.1993.1010

Kleidon, A. (2012). How does the Earth system generate and maintain thermodynamic disequilibrium and what does it imply for the future of the planet? *Philosophical Transactions of the Royal Society A: Mathematical, Physical and Engineering Sciences*, *370*(1962), 1012–1040. https://doi.org/10.1098/rsta.2011.0316

Kleidon, A., & Lorenz, R. D. (2004). *Non-equilibrium Thermodynamics and the Production of Entropy: Life, Earth, and Beyond*. Springer Science & Business Media.

Komacek, T. D., Fauchez, T. J., Wolf, E. T., & Abbot, D. S. (2020). Clouds will Likely Prevent the Detection of Water Vapor in JWST Transmission Spectra of Terrestrial Exoplanets. *The Astrophysical Journal*, *888*(2), L20. https://doi.org/10.3847/2041-8213/ab6200

Kopparapu, R. Kumar, Wolf, E. T., Arney, G., Batalha, N. E., Haqq-Misra, J., Grimm, S. L., & Heng, K. (2017). Habitable Moist Atmospheres on Terrestrial Planets near the Inner Edge of the Habitable Zone around M Dwarfs. *The Astrophysical Journal*, *845*(1), 5. https://doi.org/10.3847/1538-4357/aa7cf9

Kopparapu, R. K., Ramirez, R., Kasting, J. F., Eymet, V., Robinson, T. D., Mahadevan, S., Terrien, R. C., Domagal-Goldman, S., Meadows, V., & Deshpande, R. (2013). Habitable Zones around Main-sequence Stars: New Estimates. *The Astrophysical Journal*, *765*, 131. https://doi.org/10.1088/0004-637X/765/2/131

Kopparapu, R. K., Ramirez, R. M., SchottelKotte, J., Kasting, J. F., Domagal-Goldman, S., & Eymet, V. (2014). Habitable Zones around Main-sequence Stars: Dependence on Planetary Mass. *The Astrophysical Journal Letters*, *787*, L29. https://doi.org/10.1088/2041-8205/787/2/L29

Kuhn, T., Cunze, S., Kochmann, J., & Klimpel, S. (2016). Environmental variables and definitive host distribution: A habitat suitability modelling for endohelminth parasites in the marine realm. *Scientific Reports*, *6*(1), 30246. https://doi.org/10.1038/srep30246

Lauber, C. L., Strickland, M. S., Bradford, M. A., & Fierer, N. (2008). The influence of soil properties on the structure of bacterial and fungal communities across land-use types. *Soil Biology and Biochemistry*, *40*(9), 2407–2415. https://doi.org/10.1016/j.soilbio.2008.05.021

Lenton, T. M. (1998). Gaia and natural selection. *Nature*, *394*(6692), 439–447. https://doi.org/10.1038/28792

Lenton, T. M., & Lovelock, J. E. (2000). Daisyworld is Darwinian: Constraints on Adaptation are Important for Planetary Self-Regulation. *Journal of Theoretical Biology*, *206*(1), 109–114. https://doi.org/10.1006/jtbi.2000.2105

Lollar, G. S., Warr, O., Telling, J., Osburn, M. R., & Lollar, B. S. (2019). 'Follow the Water': Hydrogeochemical Constraints on Microbial Investigations 2.4 km Below Surface at the Kidd Creek Deep Fluid and Deep Life Observatory. *Geomicrobiology Journal*, *36*(10), 859–872. https://doi.org/10.1080/01490451.2019.1641770

Lorenz, R. D. (2020). Maunder's Work on Planetary Habitability in 1913: Early Use of the term "Habitable Zone" and a "Drake Equation" Calculation. *Research Notes of the AAS*, *4*(6), 79. https://doi.org/10.3847/2515-5172/ab9831

Lortie, C. J., Braun, J., Filazzola, A., & Miguel, F. (2020). A checklist for choosing between R packages in ecology and evolution. *Ecology and Evolution*, *10*(3), 1098–1105. https://doi.org/10.1002/ece3.5970

Lovelock, J. E. (1965). A Physical Basis for Life Detection Experiments. *Nature*, 207(4997), 568–570. https://doi.org/10.1038/207568a0

Lovelock, J. E., Kaplan, I. R., & Pirie, N. W. (1975). Thermodynamics and the recognition of alien biospheres. *Proceedings of the Royal Society of London. Series B. Biological Sciences*, 189(1095), 167–181. https://doi.org/10.1098/rspb.1975.0051

Luger, R., & Barnes, R. (2015). Extreme Water Loss and Abiotic O2 Buildup on Planets Throughout the Habitable Zones of M Dwarfs. *Astrobiology*, *15*(2), 119–143. https://doi.org/10.1089/ast.2014.1231

Lynch, M. D. J., & Neufeld, J. D. (2015). Ecology and exploration of the rare biosphere. *Nature Reviews Microbiology*, *13*(4), 217–229. https://doi.org/10.1038/nrmicro3400

Martinez-Frias, J., Lázaro, E., & Esteve-Núñez, A. (2007). Geomarkers versus Biomarkers: Paleoenvironmental and Astrobiological Significance. *AMBIO: A Journal of the Human Environment*, *36*(5), 425–426. https://doi.org/10.1579/0044-7447(2007)36[425:GVBPAA]2.0.CO;2

Martiny, J. B. H., Bohannan, B. J. M., Brown, J. H., Colwell, R. K., Fuhrman, J. A., Green, J. L., Horner-Devine, M. C., Kane, M., Krumins, J. A., Kuske, C. R., Morin, P. J., Naeem, S., Øvreås, L., Reysenbach, A.-L., Smith, V. H., & Staley, J. T. (2006). Microbial biogeography: Putting microorganisms on the map. *Nature Reviews Microbiology*, *4*(2), 102–112. https://doi.org/10.1038/nrmicro1341

Maunder, E. W. (1913). Are the planets inhabited? *London, New York, Harper & Brothers, 1913*. http://adsabs.harvard.edu/abs/1913api..book.....M

*Habitability Models* 18

McKay, C. P. (2014). Requirements and limits for life in the context of exoplanets. *Proceedings of the National Academy of Sciences*, *111*(35), 12628–12633. https://doi.org/10.1073/pnas.1304212111

Méndez, A., Schulze-Makuch, D., Nery, G., Rivera-Valentin, E. G., Davila, A., Ramirez, R., Wood, T., Rodriguez-Garcia, A. D., Soto-Soto, A., Gonzalez-Villanueva, S. E., Rivera-Saavedra, S. M., Maldonado-Vazquez, G. J., Colon-Acosta, E., Cruz-Mendoza, V. M., Crespo Sanchez, J. K., & Estevez-Mesa, N. (2018). *A General Mass-Energy Habitability Model*. *49*, 2511.

Molina, R. D., Salazar, J. F., Martínez, J. A., Villegas, J. C., & Arias, P. A. (2019). Forest-Induced Exponential Growth of Precipitation Along Climatological Wind Streamlines Over the Amazon. *Journal of Geophysical Research: Atmospheres*, *124*(5), 2589–2599. https://doi.org/10.1029/2018JD029534

National Research Council. (2007). *The Limits of Organic Life in Planetary Systems*. https://doi.org/10.17226/11919

Nowajewski, P., Rojas, M., Rojo, P., & Kimeswenger, S. (2018). Atmospheric dynamics and habitability range in Earth-like aquaplanets obliquity simulations. *Icarus*, *305*, 84–90. https://doi.org/10.1016/j.icarus.2018.01.002

Oren, A. (1999). Bioenergetic Aspects of Halophilism. *Microbiology and Molecular Biology Reviews*, *63*(2), 334–348. https://doi.org/10.1128/MMBR.63.2.334-348.1999

Oren, A. (2001). The bioenergetic basis for the decrease in metabolic diversity at increasing salt concentrations: Implications for the functioning of salt lake ecosystems. In J. M. Melack, R. Jellison, & D. B. Herbst (Eds.), *Saline Lakes: Publications from the 7th International Conference on Salt Lakes, held in Death Valley National Park, California, U.S.A., September 1999* (pp. 61–72). Springer Netherlands. https://doi.org/10.1007/978-94-017-2934-5_6

Pan, Y., Cheong, C. M., & Blevis, E. (2010). The climate change habitability index. *Interactions*, *17*(6), 29–33. https://doi.org/10.1145/1865245.1865253

Pikuta, E. V., Hoover, R. B., & Tang, J. (2007). Microbial Extremophiles at the Limits of Life. *Critical Reviews in Microbiology*, *33*(3), 183–209. https://doi.org/10.1080/10408410701451948

Radeloff, V. C., Dubinin, M., Coops, N. C., Allen, A. M., Brooks, T. M., Clayton, M. K., Costa, G. C., Graham, C. H., Helmers, D. P., Ives, A. R., Kolesov, D., Pidgeon, A. M., Rapacciuolo, G., Razenkova, E., Suttidate, N., Young, B. E., Zhu, L., & Hobi, M. L. (2019). The Dynamic Habitat Indices (DHIs) from MODIS and global biodiversity. *Remote Sensing of Environment*, *222*, 204–214. https://doi.org/10.1016/j.rse.2018.12.009

Rajakaruna, N., & Boyd, R. S. (2008). Edaphic Factor. In S. E. Jørgensen & B. D. Fath (Eds.), *Encyclopedia of Ecology* (pp. 1201–1207). Academic Press. https://doi.org/10.1016/B978-008045405-4.00484-5

Ramirez, R. M. (2018). A More Comprehensive Habitable Zone for Finding Life on Other Planets. *Geosciences*, *8*(8), 280. https://doi.org/10.3390/geosciences8080280

Ramirez, R. M. (2020). A Complex Life Habitable Zone Based On Lipid Solubility Theory. *Scientific Reports*, *10*(1), 7432. https://doi.org/10.1038/s41598-020-64436-z

Ramirez, R. M., & Kaltenegger, L. (2014). THE HABITABLE ZONES OF PRE-MAIN-SEQUENCE STARS. *The Astrophysical Journal*, *797*(2), L25. https://doi.org/10.1088/2041-8205/797/2/L25

Ramirez, R. M., & Kaltenegger, L. (2017). A Volcanic Hydrogen Habitable Zone. *The Astrophysical Journal Letters*, *837*, L4. https://doi.org/10.3847/2041-8213/aa60c8

Ramirez, R. M., & Kaltenegger, L. (2018). A Methane Extension to the Classical Habitable Zone. *The Astrophysical Journal*, *858*, 72. https://doi.org/10.3847/1538-4357/aab8fa

Rodríguez-López, L., Cardenas, R., Parra, O., González-Rodríguez, L., Martin, O., & Urrutia, R. (2019). On the quantification of habitability: Merging the astrobiological and ecological schools. *International Journal of Astrobiology*, *18*(5), 412–415. https://doi.org/10.1017/S1473550418000344

Roloff, G. J., & Kernohan, B. J. (1999). Evaluating Reliability of Habitat Suitability Index Models. *Wildlife Society Bulletin (1973-2006)*, *27*(4), 973–985. JSTOR.

Salazar, J. F., & Poveda, G. (2009). Role of a simplified hydrological cycle and clouds in regulating the climate–biota system of Daisyworld. *Tellus B: Chemical and Physical Meteorology*, *61*(2), 483–497. https://doi.org/10.1111/j.1600-0889.2009.00411.x

Schramski, J. R., Dell, A. I., Grady, J. M., Sibly, R. M., & Brown, J. H. (2015). Metabolic theory predicts whole-ecosystem properties. *Proceedings of the National Academy of Sciences of the United States of America*, *112*(8), 2617–2622. https://doi.org/10.1073/pnas.1423502112

Schulze-Makuch, D., Méndez, A., Fairén, A. G., von Paris, P., Turse, C., Boyer, G., Davila, A. F., António, M. R. de S., Catling, D., & Irwin, L. N. (2011). A Two-Tiered Approach to Assessing the Habitability of Exoplanets. *Astrobiology*, *11*(10), 1041–1052. https://doi.org/10.1089/ast.2010.0592

Schulze-Makuch, D., Heller, R., & Guinan, E. (2020). In Search for a Planet Better than Earth: Top Contenders for a Superhabitable World. *Astrobiology*, *20*(12), 1394–1404. https://doi.org/10.1089/ast.2019.2161

Seales, J., & Lenardic, A. (2020). Uncertainty Quantification in Planetary Thermal History Models: Implications for Hypotheses Discrimination and Habitability Modeling. *The Astrophysical Journal*, *893*(2), 114. https://doi.org/10.3847/1538-4357/ab822b

Selsis, F., Kasting, J. F., Levrard, B., Paillet, J., Ribas, I., & Delfosse, X. (2007). Habitable planets around the star Gliese 581? *Astronomy and Astrophysics*, *476*, 1373–1387. https://doi.org/10.1051/0004-6361:20078091

Shock, E. L., & Holland, M. E. (2007). Quantitative Habitability. *Astrobiology*, *7*(6), 839–851. https://doi.org/10.1089/ast.2007.0137

Sholes, S. F., Krissansen-Totton, J., & Catling, D. C. (2019). A Maximum Subsurface Biomass on Mars from Untapped Free Energy: CO and H2 as Potential Antibiosignatures. *Astrobiology*, *19*, 655–668. https://doi.org/10.1089/ast.2018.1835

Silva, L., Vladilo, G., Schulte, P. M., Murante, G., & Provenzale, A. (2017). From climate models to planetary habitability: Temperature constraints for complex life. *International Journal of Astrobiology*, *16*(3), 244–265. https://doi.org/10.1017/S1473550416000215

Spiegel, D. S., Menou, K., & Scharf, C. A. (2009). HABITABLE CLIMATES: THE INFLUENCE OF OBLIQUITY. *The Astrophysical Journal*, *691*(1), 596–610. https://doi.org/10.1088/0004-637X/691/1/596

Stoker, C. R., Zent, A., Catling, D. C., Douglas, S., Marshall, J. R., Archer, D., Clark, B., Kounaves, S. P., Lemmon, M. T., Quinn, R., Renno, N., Smith, P. H., & Young, S. M. M. (2010). Habitability of the Phoenix landing site. *Journal of Geophysical Research: Planets*, *115*(E6). https://doi.org/10.1029/2009JE003421

Taubner, R.-S., Olsson-Francis, K., Vance, S. D., Ramkissoon, N. K., Postberg, F., de Vera, J.-P., Antunes, A., Camprubi Casas, E., Sekine, Y., Noack, L., Barge, L., Goodman, J., Jebbar, M., Journaux, B., Karatekin, Ö., Klenner, F., Rabbow, E., Rettberg, P., Rückriemen-Bez, T., … Soderlund, K. M. (2020). Experimental and Simulation Efforts in the Astrobiological Exploration of Exooceans. *Space Science Reviews*, *216*(1), 9. https://doi.org/10.1007/s11214-020-0635-5

Tecon, R., & Or, D. (2017). Biophysical processes supporting the diversity of microbial life in soil. *FEMS Microbiology Reviews*, *41*(5), 599–623. https://doi.org/10.1093/femsre/fux039

Treseder, K. K., Balser, T. C., Bradford, M. A., Brodie, E. L., Dubinsky, E. A., Eviner, V. T., Hofmockel, K. S., Lennon, J. T., Levine, U. Y., MacGregor, B. J., Pett-Ridge, J., & Waldrop, M. P. (2012). Integrating microbial ecology into ecosystem models: Challenges and priorities. *Biogeochemistry*, *109*(1), 7–18. https://doi.org/10.1007/s10533-011-9636-5

Krissansen-Totton, J., Bergsman, D. S., & Catling, D. C. (2016). On Detecting Biospheres from Chemical Thermodynamic Disequilibrium in Planetary Atmospheres. *Astrobiology*, 16(1), 39–67. https://doi.org/10.1089/ast.2015.1327

U. S. Fish and Wildlife Service Division of Ecological. (1980). *Ecological services manual*. Division of Ecological Services, U.S. Fish and Wildlife Service, Department of the Interior.

Underwood, D. R., Jones, B. W., & Sleep, P. N. (2003). The evolution of habitable zones during stellar lifetimes and its implications on the search for extraterrestrial life. *International Journal of Astrobiology*, *2*, 289–299. https://doi.org/10.1017/S1473550404001715

Van Beusekom, A. E., González, G., & Scholl, M. A. (2017). Analyzing cloud base at local and regional scales to understand tropical montane cloud forest vulnerability to climate change. *Atmospheric Chemistry and Physics*, *17*(11), 7245–7259. https://doi.org/10.5194/acp-17-7245-2017

Van der Meer, J. (2006). Metabolic theories in ecology. *Trends in Ecology & Evolution*, *21*(3), 136–140. https://doi.org/10.1016/j.tree.2005.11.004

Walker, J. C. G., Hays, P. B., & Kasting, J. F. (1981). A negative feedback mechanism for the long-term stabilization of Earth's surface temperature. *Journal of Geophysical Research: Oceans*, *86*(C10), 9776–9782. https://doi.org/10.1029/JC086iC10p09776

Williams, J. P., & Hallsworth, J. E. (2009). Limits of life in hostile environments: No barriers to biosphere function? *Environmental Microbiology*, *11*(12), 3292–3308. https://doi.org/10.1111/j.1462-2920.2009.02079.x

Wright & Gelino (eds.). (2019). NASA and the Search for Technosignatures: A Report from the NASA Technosignatures Workshop. *ArXiv:1812.08681 [Astro-Ph, Physics:Physics]*. http://arxiv.org/abs/1812.08681

Yin, X., Kropff, M. J., McLaren, G., & Visperas, R. M. (1995). A nonlinear model for crop development as a function of temperature. *Agricultural and Forest Meteorology*, *77*(1), 1–16. https://doi.org/10.1016/0168-1923(95)02236-Q

Zaks, D. P. M., Ramankutty, N., Barford, C. C., & Foley, J. A. (2007). From Miami to Madison: Investigating the relationship between climate and terrestrial net primary production. *Global Biogeochemical Cycles*, *21*(3). https://doi.org/10.1029/2006GB002705

Zsom, A. (2015). A POPULATION-BASED HABITABLE ZONE PERSPECTIVE. *The Astrophysical Journal*, *813*(1), 9. https://doi.org/10.1088/0004-637X/813/1/9

Zuluaga, J. I., Salazar, J. F., Cuartas-Restrepo, P., & Poveda, G. (2014). The Habitable Zone of Inhabited Planets. *Biogeosciences Discussions*, *11*(6), 8443–8483. https://doi.org/10.5194/bgd-11-8443-2014